\documentclass[]{spie}  

\usepackage{amsmath,amsfonts,amssymb}
\usepackage{CJK}
\usepackage{graphicx}
\usepackage[colorlinks=true, allcolors=blue]{hyperref}
\usepackage{multirow}
\usepackage{booktabs}
\usepackage{graphicx}

\title{Robo-AO-2: entering the era of automated science operations, hybrid wavefront sensing, and adaptive secondary integration}

\author[a]{Christoph Baranec*}
\author[b]{Reed Riddle}
\author[a]{James Ou}
\author[a]{Ruihan Zhang}
\author[a]{Guillaume Huber}
\author[a]{Zachary Werber}
\author[a]{Luke Mckay}
\author[a]{Rachel Rampy}
\author[a]{Michael C. Liu}
\author[c]{Carl Ziegler}
\author[a]{Mark Chun}
\author[$\!$]{Keith Powell}
\author[d]{Marcos A. van Dam}

\affil[a]{Institute for Astronomy, University of Hawaii at Manoa, Hilo, HI 96720 USA}
\affil[b]{California Institute of Technology, Pasadena, CA 91125 USA}
\affil[c]{Department of Physics, Engineering and Astronomy, Stephen F. Austin State University, TX 75962, USA}
\affil[d]{Flat Wavefronts, Christchurch 8022, New Zealand}

\authorinfo{* baranec@hawaii.edu; phone 1-808-498-9817; http://robo-ao.org/}

\begin{document} 
\maketitle

\begin{abstract}
We present the first science results and new technical milestones from the Robo-AO-2 facility at the University of Hawaii 2.2-m telescope. Following successful commissioning, the system began science operations in 2025. We are starting a large-scale survey of  young stars in the Scorpius-Centaurus association to detect sub-stellar companions, vetting the Habitable Worlds Observatory Target Stars and Systems list, and discovering stellar blends for TESS exoplanet host candidates. We report on the commissioning of the natural guide star wavefront sensor, supporting science and future hybrid laser-stellar sensing demonstrations. Finally, we detail the automation of the telescope facility, Robo88, via updated telescope control systems and absolute encoders, and the integration of Robo-AO-2 with the telescope's new adaptive secondary mirror.
\end{abstract}


\section{INTRODUCTION}
\label{sec:intro}  

Robo-AO-2 is a robotic laser adaptive optics system built by the Institute for Astronomy for the University of Hawaii 2.2-m telescope (UH2.2m) on Maunakea, Hawaii\cite{Baranec2024}. Robo-AO-2 comprises a UV laser projector, a bent-Cassegrain mounted adaptive optics system with low-noise, high-speed visible and infrared imaging arrays that can double as tip-tilt sensors, a reconfigurable stellar wavefront sensor, and a set of electronics and computers with additional functionality. It is based on the prototype Robo-AO system\cite{Baranec2013, Baranec2014, Baranec2021} which was used at the Palomar 1.5-m telescope\cite{cenko} (2011-2015), the 2.1-m telescope at Kitt Peak \cite{RAO_KP} (2015-2018) and then at the UH2.2m \cite{LASSO} (2019-2020). The core innovation with the Robo-AO platform is the full automation of the adaptive optics system, science instruments and data reduction. When paired with a similarly automated telescope, this combination allows for very efficient survey, monitoring, and target-of-opportunity observing. 

\section{Initial Scientific results}
\label{sec:science}

Robo-AO-2 began commissioning in May 2023, with early imaging capabilities demonstrated in Baranec, et al. 2024\cite{Baranec2024}. In 2025, we began shared-risk science. 

\subsection{Binary star relative astrometry}
\label{sec:binary}

\begin{figure} [ht]
   \begin{center}
   \includegraphics[width=\textwidth]{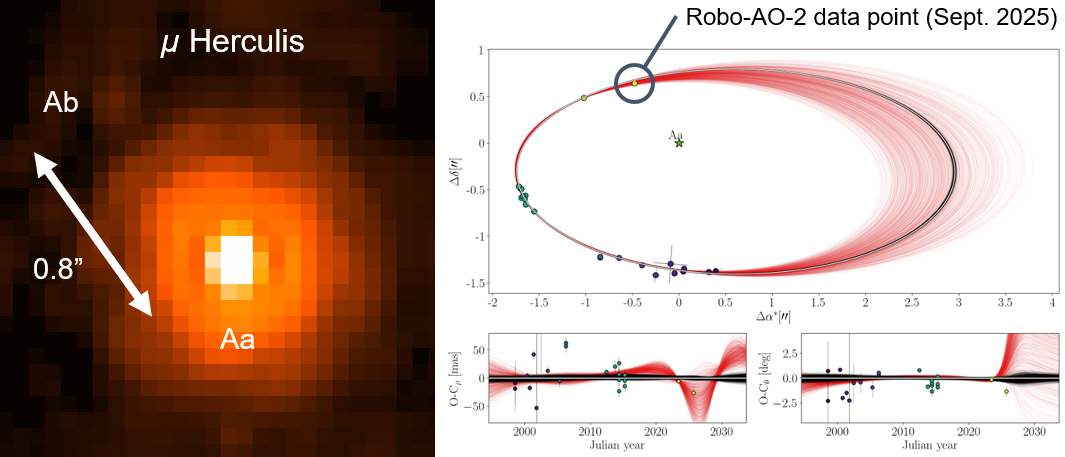}
   \end{center}
   \caption[example] 
   { \label{fig:1} 
\textit{Left:} H-band image of $\mu$ Herculis Aa and Ab. North is up, East is right. \textit{Right:} Relative astrometric location of the two components over nearly 3 decades. Adapted from Figure 3 of Marcussen, et al. 2026.\cite{muHer} }
   \end{figure}

In August 2025, one of our collaborators had Keck/NIRC2 time to observe the Aa-Ab components of the nearby quadruple star system $\mu$ Herculis. Unfortunately, their observing run was canceled due to dome shutter issues, so they appealed to us to acquire the necessary images. We observed $\mu$ Herculis on September 17 2025 using the natural guide star wavefront sensor to drive the adaptive optics correction as it has an apparent V magnitude of 3.4.  The target was observed in the H band for a total of 5 minutes using 3 ms exposures. Each exposure was corrected for detector bias and sky-background before sorting by image quality. Using the best 10\% of frames, aligned to the image position of Aa, we created a final co-added image (see Fig.~\ref{fig:1}).

To calibrate the astrometry of the Robo-AO-2 instrument, we observed several close binary stars from the OK Binary Star Catalog\cite{OKB}. We determined stellar locations using the Aperture Photometry Tool and calculated the pixel scale on the infrared camera to be 65.5$\pm$0.1 mas/pixel with a position angle offset from up to North of $-0.4^\circ\pm0.1^\circ$. We measured the $\mu$ Herculis Aa and Ab pair to be separated by 814$\pm$13 mas with a position angle of $323.6^\circ\pm0.5^\circ$.

\begin{figure} [h]
   \begin{center}
   \includegraphics[width=5.5in]{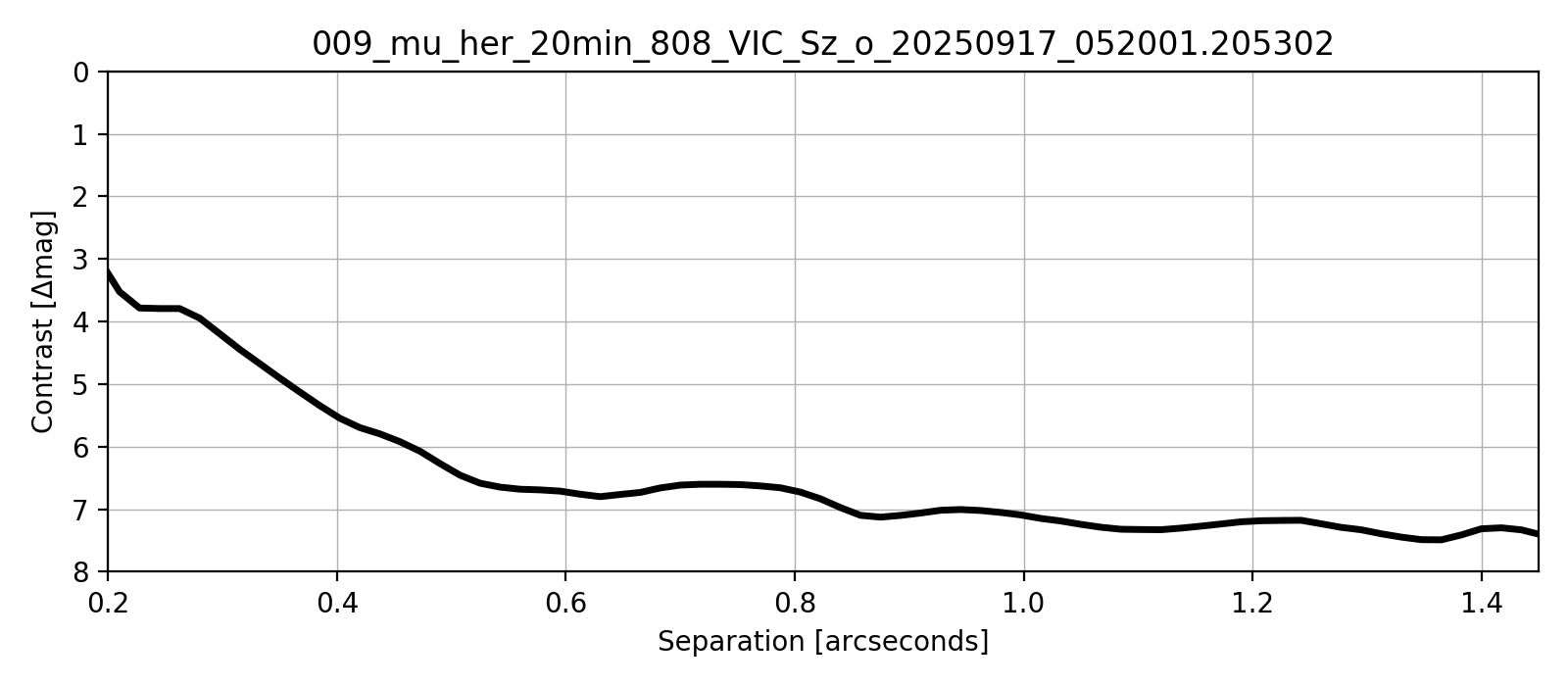}
   \end{center}
   \caption[example] 
   { \label{fig:2} 
Calculated 5-$\sigma$ contrast curve for $\mu$ Herculis at $\overline{\lambda} = 808$nm. Details on the contrast curve generation are found in Jensen-Clem, et al., 2018\cite{RAO_KP}. }
   \end{figure}

In addition to the infrared observation, we also observed $\mu$ Herculis for 20 minutes with the Robo-AO-2 visible camera using a narrow-band filter: $\overline{\lambda} = 808$ nm, $\Delta\lambda = 5$ nm. We were unable to detect Ab at the 5-$\sigma$ level and determined that Ab is at least 6.5 magnitudes fainter than Aa at this wavelength (see Fig.~\ref{fig:2}).

\subsection{Robotic AO Survey of Sco-Cen for Companions at Large Separations (RASSCALS)}

The direct discovery and characterization of giant planets and brown dwarfs is one of the major frontiers in astronomy. Direct imaging has revealed a small population of young companions with masses spanning the planetary and brown dwarf regimes, providing unique laboratories for studying atmospheric physics, formation pathways, and the architecture of planetary systems. However, such companions are intrinsically rare.  Previous adaptive optics imaging searches for substellar companions to young stars have typically observed  a few hundred stars, with each survey yielding only a handful of discoveries. As a result, the demographics of imaged companions remain poorly constrained.

\begin{figure} [ht]
   \begin{center}
   \includegraphics[width=\textwidth]{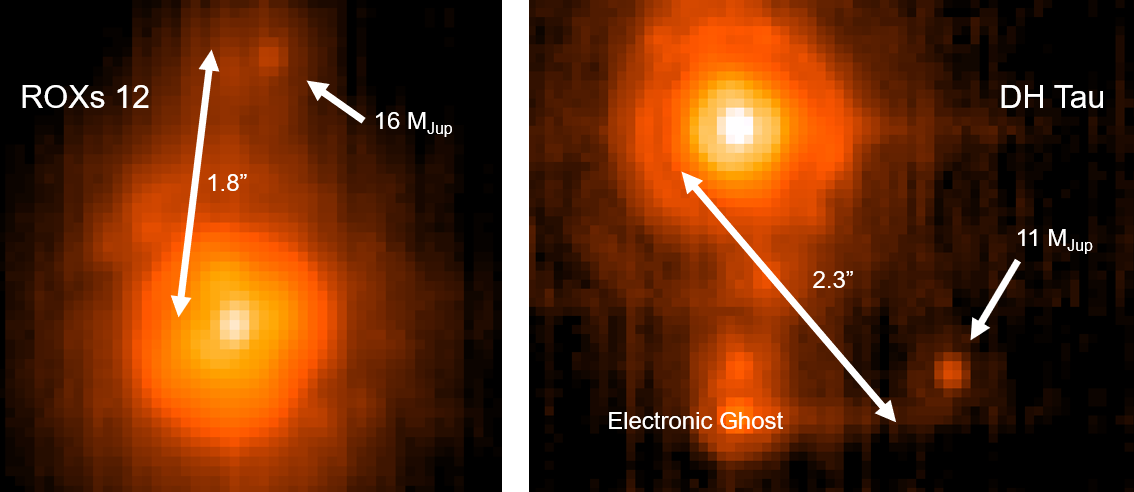}
   \end{center}
   \caption[example] 
   { \label{fig:3} 
Robo-AO-2 infrared ($H$-band) images using laser guide star correction.
\textit{Left:} ROXs 12, a young star with a wide-separation $16~M_{\rm Jup}$ companion\cite{Kraus2014}. \textit{Right:} DH Tau, a young star with a wide-separation $11~M_{\rm Jup}$ companion\cite{Itoh2005}. The electronic ghost at 6 o’clock is consistent and thus can be model-subtracted with only small impact on sky coverage.}
   \end{figure}

\begin{figure} [ht]
   \begin{center}
   \includegraphics[width=\textwidth]{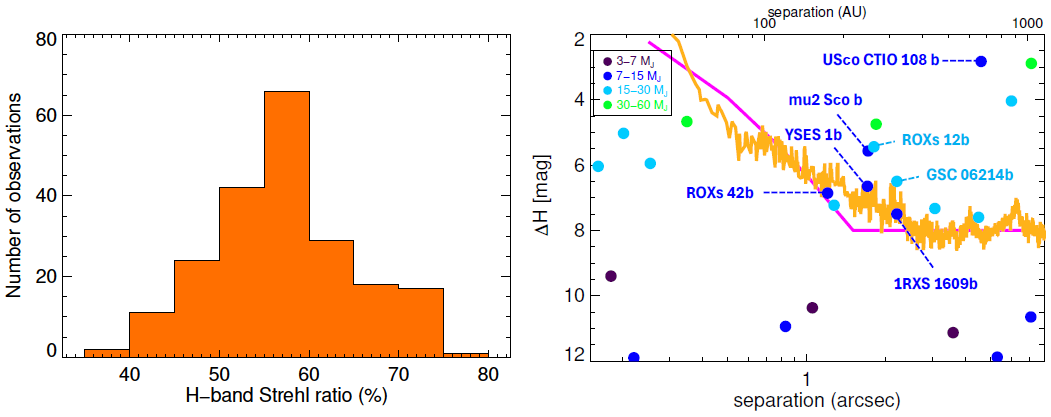}
   \end{center}
   \caption[example] 
   { \label{fig:4} 
\textit{Left:} Robo-AO-2 delivers excellent $H$-band images, with an expected median Strehl ratio of 57\%. The histogram summarizes 210 $z^\prime$-band measurements taken in various weather conditions during 2025B commissioning and early science for another program, using the Mar\'{e}chal approximation to convert from $z^\prime$ to $H$. \textit{Right}: Predicted (magenta) and measured (gold) $H$-band 5-$\sigma$ contrast curve for Robo-AO-2 compared to known Sco-Cen substellar companions from The UltracoolSheet (colored points). Objects above the contrast curve can be detected. Mass estimates for the companions are from the literature, with a few notable companions labeled (e.g., 1RXS 1609b was one of the first directly imaged planets\cite{L2008, L2010}). A representative distance of 145 pc is assumed for the top axis.}
   \end{figure}

One of our first major science programs is the Robo-AO Survey of Sco-Cen for Large-Separation Companions (RASSCALS). The survey targets $\sim$2000 young stars in the Scorpius--Centaurus OB association, the nearest large star-forming complex to the Sun. Given its age of $\sim$10--20 Myr, giant planets and brown dwarfs remain self-luminous and are readily detectable at near-infrared wavelengths. Robo-AO-2 is expected to achieve sensitivity to companions with masses greater than approximately 5 $M_{\rm Jup}$ at projected separations of hundreds of astronomical units (see Figs.~\ref{fig:3} and \ref{fig:4}).

The survey occupies a complementary niche relative to previous direct-imaging campaigns. Extreme-AO surveys such as Gemini/GPIES\cite{Nealson2019} and VLT/SHINE\cite{Vegan2021} focused on achieving the deepest possible contrasts at small separations, probing young giant planets at tens of AU. Robo-AO-2 instead targets a much larger stellar sample while remaining sensitive to a region of parameter space already known to contain planetary-mass and brown dwarf companions. The goal is not necessarily to discover the lowest-mass or closest-in planets, but rather to substantially increase the census of wide companions through a homogeneous survey of a single young stellar population.

The scientific motivation extends beyond simple discovery statistics. Existing occurrence-rate estimates for wide-orbit companions are derived from relatively small samples and are therefore subject to large uncertainties. Current studies suggest that planetary-mass companions and brown dwarfs occupy overlapping regions of mass and separation space, raising fundamental questions about whether they arise from distinct formation mechanisms or represent a continuous population. A substantially larger sample is needed to determine the underlying distributions of companion mass, orbital separation, and host-star properties.

Simulations based on published companion occurrence rates indicate that a survey of this scale could discover several new planetary-mass companions and several dozen brown dwarfs. Such a sample would represent one of the largest homogeneous collections of directly imaged wide-orbit substellar companions assembled to date. The resulting discoveries would enable statistically meaningful investigations of companion demographics, including occurrence rates, mass functions, and separation distributions.  Especially notable will be the fact that all the companions will reside in the same star-forming complex, in contrast to the current census of substellar companions which reside around young stars belonging to  different stellar associations and the field population as well.

Robo-AO-2 surveys are also highly complementary to upcoming astrometric discoveries from Gaia Data Release 4. Gaia will be most sensitive to companions at separations of a few to tens of AU, while direct imaging is most effective for wider companions. Together, these techniques provide coverage across a broad range of orbital separations and offer a more complete picture of giant-planet architectures.

Newly discovered companions will become valuable targets for detailed characterization with larger facilities. Because Robo-AO-2 is sensitive to relatively bright, widely separated companions, follow-up observations with 8--10 m telescopes and space observatories can readily obtain photometry and spectroscopy. These measurements will constrain atmospheric properties, surface gravities, accretion signatures, and elemental abundances, providing insights into the formation and evolution of giant planets and brown dwarfs.

This science case highlights the broader philosophy of Robo-AO-2. By combining adaptive optics image quality with robotic observing efficiency, the instrument enables surveys that would be prohibitively expensive on traditional AO facilities. The resulting large, homogeneous datasets provide a powerful complement to targeted observations with flagship observatories and establish Robo-AO-2 as an important discovery engine for a new large sample of exoplanet and substellar companion studies.

\subsection{Resolving the Stellar Environments of TESS Planet Hosts}

The Transiting Exoplanet Survey Satellite (TESS) has transformed exoplanet discovery by identifying thousands of planetary candidates around bright nearby stars. However, a major challenge in interpreting these discoveries is the presence of unresolved stellar companions. Close binary stars can mimic planetary signals, alter inferred planetary properties through photometric dilution, and complicate follow-up characterization efforts. High-angular-resolution imaging is therefore an essential component of exoplanet validation and characterization programs.

Robo-AO-2 provides a unique capability for addressing this challenge through simultaneous visible and near-infrared adaptive optics imaging combined with highly automated operations. Unlike conventional adaptive optics facilities, which typically observe a limited number of targets per night, Robo-AO-2 can efficiently survey hundreds of stars in a single observing session. This combination of diffraction-limited imaging and survey-scale throughput enables a comprehensive characterization of the stellar environments surrounding TESS planet candidates.

One of the first large-scale science programs planned for Robo-AO-2 is a survey of approximately 4,000 TESS Objects of Interest (TOIs) observable from Maunakea. The survey is designed to provide a nearly complete census of close stellar companions around TESS planet hosts while simultaneously filling gaps left by previous speckle and adaptive optics surveys. Robo-AO-2's simultaneous dual-band imaging is particularly valuable because visible observations closely match the TESS bandpass, while near-infrared observations provide enhanced sensitivity to low-mass stellar companions. Together, these measurements enable improved determinations of companion properties and assessments of physical association.

\begin{figure} [ht]
   \begin{center}
   \includegraphics[width=300pt]{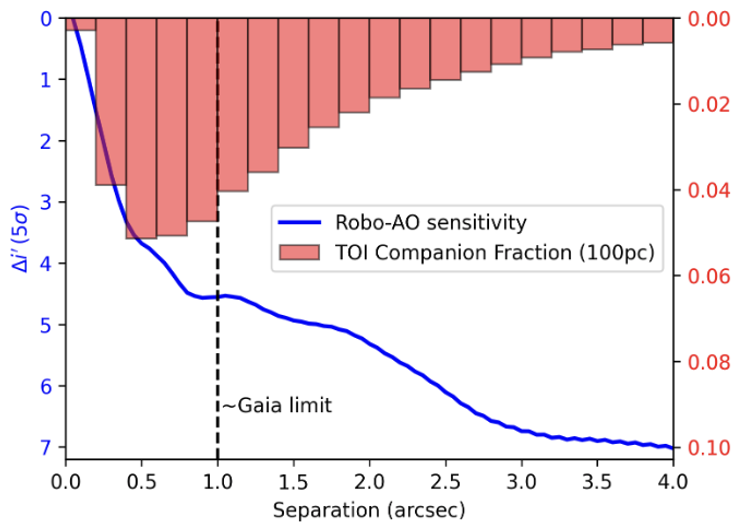}
   \end{center}
   \caption[example] 
   { \label{fig:7} 
Robo-AO-2 probes the critical sub-arcsecond regime where most bound companions reside. Visible contrast curve ($\Delta$i$^\prime$) overlaid on the expected companion distribution (red histogram) for a typical TOI at 100 pc, generated from field binary statistics\cite{Raghavan2010} and truncated at low separations as found for binary planet hosts\cite{Kraus2016}. Robo-AO-2 detects the vast majority of bound companions, critically covering the peak of the distribution at $< 1''$ separations where Gaia is largely insensitive.\cite{Ziegler2018}}
   \end{figure}

The survey targets a separation regime that is difficult to probe with existing all-sky astrometric surveys. Most stellar companions reside within projected separations corresponding to less than one arcsecond for typical TESS targets, where catalog completeness rapidly declines. Robo-AO-2 routinely achieves angular resolutions of approximately 80--90 mas at red wavelengths, allowing the detection and characterization of companions throughout the critical sub-arcsecond region (see Fig.~\ref{fig:7}). Near-infrared observations further improve the sensitivity to faint late-type stars and low-mass companions, which can significantly influence the interpretation of planetary systems.

Beyond identifying previously unresolved companions, Robo-AO-2 enables multi-epoch monitoring of binary systems. Repeated observations over timescales of several years can reveal orbital motion, confirm physical association, and provide constraints on orbital architecture. Early demonstrations of Robo-AO-2 have already shown the ability to measure significant orbital motion in close stellar systems while simultaneously detecting faint substellar companions. Extending this capability to thousands of TESS targets will create one of the largest homogeneous datasets of planetary systems in binary environments.

The resulting survey will provide important insights into the relationship between stellar multiplicity and planet formation. Previous studies have suggested that close stellar companions suppress planet occurrence and may alter the distribution of planetary radii and orbital architectures. However, most existing investigations have relied on relatively limited samples concentrated in specific stellar populations. By leveraging the all-sky nature of the TESS mission and the high observing efficiency of Robo-AO-2, this survey will extend such studies to a much broader range of stellar masses, ages, and Galactic environments.

M-dwarf planetary systems represent a particularly compelling target for Robo-AO-2. These stars host many of the small planets most amenable to atmospheric characterization and habitability studies, yet their multiplicity properties remain incompletely explored. The near-infrared sensitivity of Robo-AO-2 is well matched to detecting low-mass companions around these cool stars, enabling a systematic assessment of how stellar multiplicity influences the formation and evolution of terrestrial planets.


\subsection{Stellar Multiplicity Survey of Habitable Worlds Observatory Targets}
\label{sec:hwo_survey}

The Habitable Worlds Observatory (HWO) is a flagship mission concept aimed at surveying approximately 25 nearby Earth-like exoplanets to search for atmospheric biosignatures and conduct transformative general astrophysics. To optimize the observatory's design and enhance its scientific return, a rigorous characterization of the host stars in the potential target catalog is required. In particular, stellar multiplicity is a critical parameter: close binary companions (separation $< 10''$) present substantial challenges for starlight suppression systems, as standard coronagraphs are typically optimized to suppress a single star. Furthermore, the statistics of companion distributions directly influence broader mission design trade-offs, such as the primary mirror aperture and coronagraph selection, while significantly altering our understanding of planetary stability and habitability in multi-star systems.

To address this precursor science requirement, we are conducting a high-angular-resolution direct-imaging survey of the HWO Tier~2 target list using Robo-AO-2. The Tier~2 list represents a high-yield sub-sample of 659 nearby, bright stars drawn from the Habitable Worlds Observatory Planet Candidate (HPIC) catalog and the Exoplanet Exploration Program (ExEP) star list, filtered by exoplanet yield modeling. Of these, 508 targets are visible from Maunakea, spanning a stellar population dominated by G-type ($38.8\%$), F-type ($31.1\%$), and K-type ($13.6\%$) stars, with smaller fractions of A-type ($13\%$) and M-type ($3.5\%$) stars. All targets are located within the local solar neighborhood ($d < 40$~pc) and are exceptionally bright ($V < 10$~mag), making them ideal candidates for natural guide star adaptive optics correction.

\begin{table}[ht]
\centering
\caption{Stellar type distribution of the 508 Maunakea-accessible HWO Tier 2 targets surveyed by Robo-AO-2.}
\label{tab:stellar_types}
\begin{tabular}{lcccccc}
\hline
\textbf{Stellar Type} & \textbf{A} & \textbf{F} & \textbf{G} & \textbf{K} & \textbf{M} & \textbf{Total} \\ \hline
Number of Stars       & 66         & 158        & 197        & 69         & 18         & 508            \\
Fraction (\%)         & 13.0\%     & 31.1\%     & 38.8\%     & 13.6\%     & 3.5\%      & 100\%          \\ \hline
\end{tabular}
\end{table}

This survey also builds on a historical baseline: 248 of these 508 stars were previously observed during 2012--2013 with the first-generation Robo-AO system\cite{Salama2022}, yielding 59 companion candidates. Re-observing these targets with Robo-AO-2 provides a valuable second epoch with a $\sim$13-year orbital baseline. Furthermore, Robo-AO-2 delivers a $\sim$2~magnitude deeper contrast ratio under median observing conditions compared to its predecessor. This enhanced sensitivity enables the detection of low-mass stellar companions down to the M-dwarf regime for the majority of the sample, and substellar brown dwarf companions for a select subset of nearby targets. Additionally, cataloging unbound background objects within the field of view is critical, as these background line-of-sight sources represent significant visual confusion and noise for future HWO coronagraphic observations. 

To date, the program has demonstrated excellent data quality and is over 80\% complete. The final high-resolution dataset will be combined with proper motion anomalies between Hipparcos and Gaia to precisely constrain binary orbital dynamics, accurately measure component masses, and refine the local mass-luminosity relation, providing a foundational database for HWO mission planning.

\section{Natural guide star wavefront sensor and hybrid AO demonstrations}

The Robo-AO 2 instrument was designed with a natural guide star wavefront sensor to support hybrid AO demonstrations (combining high-order laser and adjustable-spatial-order stellar wavefront measurements to improve stellar AO correction on fainter targets)\cite{Baranec2024, hapa} and is installed in the instrument. While similar to the already-integrated laser guide star wavefront sensor, the natural guide star wavefront sensor additionally includes swappable microlens arrays (16×16, 8×8, 5×5, 4×4, 2×2 or 1×1) on a mechanized stage, as well as a neutral-density filter wheel to artificially dim the guide star. Thus far, we have enabled the 16×16 mode in mid 2025 to support the science described in sections~\ref{sec:binary}~and~\ref{sec:hwo_survey}.  

As part of the natural guide star correction, we found the current tip-tilt mirror to be a challenge to work with, as it suffers from significant hysteresis. We recently had delivered an upgraded version of this tip-tilt mirror with a closed-loop control via strain gauge sensors and we will be installing it in Q3 2026. 
 
Currently, the laser and natural guide star wavefront sensor systems are working separately, but they will need to be working together to enable hybrid AO demonstrations. Each camera has its own frame grabber card on the computer that operates the AO loop, so we will write software to grab images from both cameras asynchronously, reconstruct both wavefronts, and implement a minimum variance tomographic wavefront reconstructor (via a vector-matrix multiplication) to update the deformable mirror position. Similar to the configuration software for the natural guide star wavefront sensor, we will enable configurations for the laser plus natural wavefront sensor in each of the microlens array modes of the natural guide star wavefront sensor.

\section{Automation and integration with Robo88}

To achieve efficient observations with Robo-AO-2, the instrument must fully control the operation of the telescope. We are currently developing the interface layer to the new telescope control system (TCS) and testing and tuning the Robo-AO-2 interface software, robotic observing sequence, and error management system\cite{Riddle2026}. Of utmost importance is ensuring the safe operation of the laser guide star on the sky to avoid striking non-natural satellites\cite{laser2026}. Although we switched to the new UH2.2m TCS in Q2 2026, we are awaiting the final installation of absolute encoders and subsequent pointing control before we start our fully automated large science surveys.

In 2019, we embarked on an effort to fully automate the UH2.2m and operation of its three instruments, called Robo88\cite{robo88}. As one of the three instruments mounted to the telescope, Robo-AO-2 will be called upon as necessary by the master observatory sequencer to observe requested targets. This will allow for potentially game-changing observing to support multi-messenger astronomy, such as taking a spectrum of a nearby supernova with the Supernova Integral Field Spectrograph and immediately thereafter taking a high-resolution image with Robo-AO-2 pinning down its exact position within its host galaxy, or simply executing diffraction-limited follow-up observations within a few minutes of detecting an event while another instrument is being used. As an initial operational mode, Robo-AO-2 will be set up to take over the telescope for blocks of time (e.g., two hours in a row during a night), and then we will develop the routines necessary to interweave observations with the other two instruments.

The automated Robo88 observatory is undergoing integration and testing with the wide field camera now, and with the spectrograph immediately thereafter. Robo-AO-2 will be the last instrument to be folded into the mix. While Robo88 will directly control the two other instruments, we will modify our automated Robo-AO-2 software to interface with the Robo88 master sequencer to allow efficient observing of the same target with multiple instruments.

\section{Integration with the UH2.2m adaptive secondary mirror}

The UH2.2m will be used to demonstrate the NSF-funded  new technology adaptive secondary mirrors\cite{nsfASM, Lee2024, Zhang2024} (ASMs) that are based on a hybrid variable reluctance actuator developed by the Nederlandse Organisatie voor Toegepast Natuurwetenschappelijk Onderzoek (TNO)\cite{Kuiper2016}. The key technological advance is that these mirrors provide more than an order of magnitude greater force output and nearly two orders of magnitude better electrical efficiency than existing voice-coil based ASMs. To characterize the UH2.2m ASM use with adaptive optics, we will integrate it with the Robo-AO-2 system. The low-noise, fast-readout natural guide star wavefront sensor camera avoids introducing focal anisoplanatism to the AO loop and offers the spatial and temporal resolution needed to drive the ASM at the UH2.2m. The visible and near-infrared science cameras within Robo-AO-2 are ideal for evaluating the performance of the ASM across a broad wavelength range. A key task is to integrate control of the ASM into Robo-AO-2's real-time controller. This involves software modifications to the real-time controller to enable communication with the drive electronics of the ASM (while simultaneously sending flat commands to the Robo-AO-2 internal MEMS deformable mirror) and to implement the proportional-integral-derivative control servo. We will then be able to quickly switch AO correction between the ASM and the well-characterized MEMS mirror, allowing us to directly compare the two.

\begin{figure} [ht]
   \begin{center}
   \includegraphics[width=\textwidth]{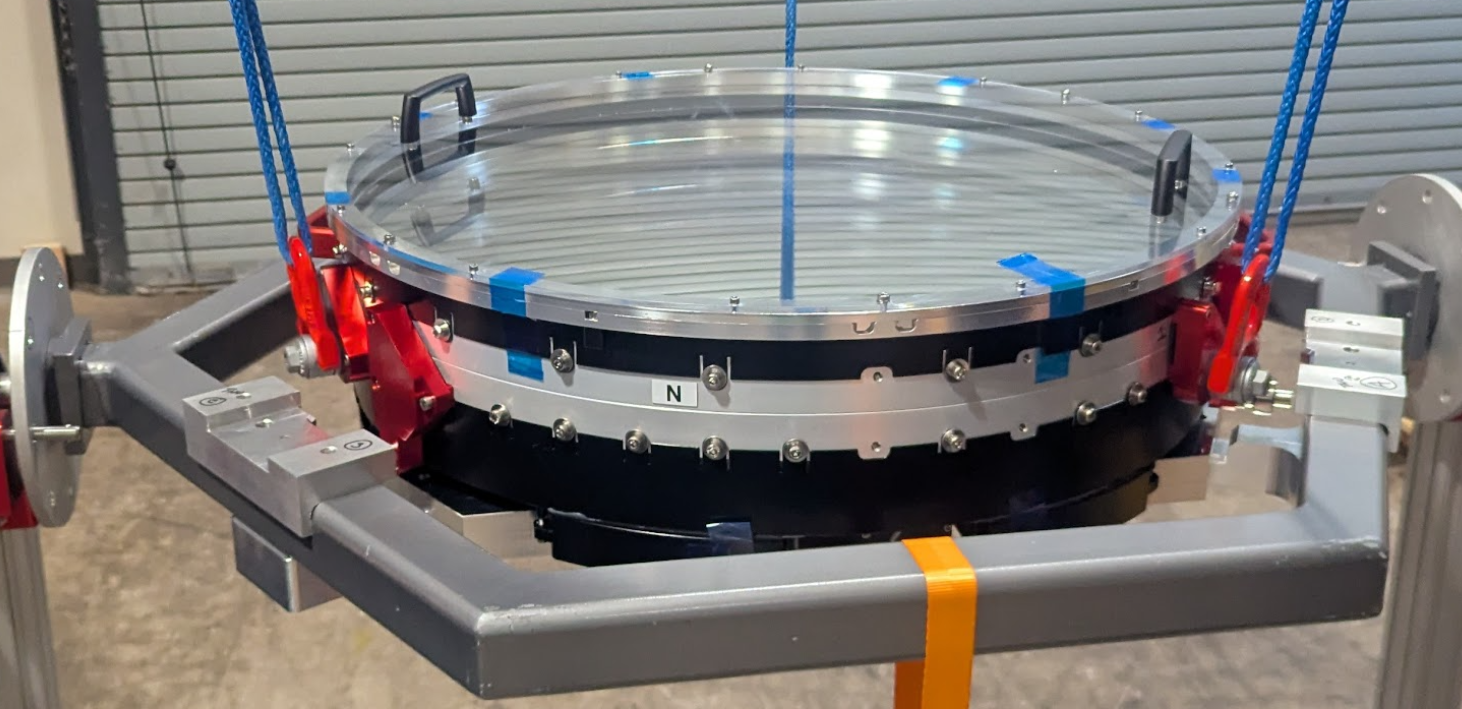}
   \end{center}
   \caption[example] 
   { \label{fig:ASM} 
The TNO adaptive secondary mirror for the University of Hawaii 2.2 m telescope currently at the Institute for Astronomy facility in Hilo, Hawaii.}
   \end{figure}

The UH2.2m ASM was delivered in June 2026 (see Fig.~\ref{fig:ASM}\cite{Kuiper2026}) and we are developing the software to control this ASM from the Robo-AO-2 system to perform our on-sky demonstration. This ASM needs to be demonstrated and evaluated on-sky as soon as possible as other major telescope facilities, e.g., Keck\cite{Lu2024}, Gemini\cite{Blain2024}, and TMT, are currently investigating this technology as an alternative to and a significant improvement over the voice-coil ASMs produced by AdOptica whose ASMs have all experienced major failures in operation\cite{Christou2014, Huerta2024}.

\acknowledgments 

The Robo-AO-2 system is supported by the National Science Foundation under Grant Nos. AST-1712014 and AST-2509941, by the State of Hawaii Capital Improvement Projects, by a gift from the Lumb Family, and by the Mt. Cuba Astronomical Foundation. Support for the infrared camera for Robo-AO and Robo-AO-2 was provided by the Mt. Cuba Astronomical Foundation and through the National Science Foundation under Grant No. AST-1106391.
 
The authors thank the people of the State of Hawaii for their support of astronomy. We are most fortunate to have the opportunity to work and conduct observations from public land on Maunakea. We are grateful to the UH2.2-m and IfA staff: Michael Yabe, Scott Caceres, Reid Ikeda, Maka Littorin, and Brian Landis.

\bibliography{Baranec_SPIE_2024} 
\bibliographystyle{spiebib} 

\end{document}